\documentclass[a4paper,11pt]{article}

\usepackage{fullpage}
\usepackage[T1]{fontenc}

\usepackage{amssymb,amsmath,cite}
\usepackage{hyperref,comment,xcolor,bbold}

\usepackage{amsthm}

\begin{document}
\vspace{5mm}
\vspace{0.5cm}

\def\thefootnote{\arabic{footnote}}
\setcounter{footnote}{0}

\allowdisplaybreaks 

\begin{titlepage}
\thispagestyle{empty}

\begin{flushright}
	\hfill{\ } 
\end{flushright}
				
\vspace{35pt}
				
\begin{center}
	{\LARGE{\bf Weak gravity versus de Sitter}}
									
	\vspace{50pt}
							
	{N.~Cribiori$^{1}$, G.~Dall'Agata$^{2,3}$ and F.~Farakos$^{2,3}$}
							
	\vspace{25pt}
							
	{
	{\it $^{1}$Institute for Theoretical Physics, TU Wien, 
	\\ Wiedner Hauptstrasse 8-10/136, A-1040 Vienna, Austria}
										
		\vspace{15pt}

		{\it  $^{2}$Dipartimento di Fisica e Astronomia ``Galileo Galilei''\\
			Universit\`a di Padova, Via Marzolo 8, 35131 Padova, Italy}
										
		\vspace{15pt}
										
		{\it   $^{3}$INFN, Sezione di Padova \\
		Via Marzolo 8, 35131 Padova, Italy}
		}
								
\vspace{40pt}
								
{ABSTRACT} 
\end{center}

We show that one can uncover a Dine--Seiberg problem for de Sitter critical points in supergravity theories by utilizing the magnetic weak gravity conjecture.
We present a large variety of N=2 gauged supergravity models that include vector multiplets and in all cases we find that the weak gravity conjecture threatens de Sitter.
A common feature in all such examples is a degenerate mass matrix for the gravitini, which we therefore deem a swampland criterion for de Sitter critical points.

\vspace{10pt}
			
\bigskip
			
\end{titlepage}

\numberwithin{equation}{section}

\baselineskip 6 mm


\section{Introduction} 

One of the most difficult problems in both minimal and extended four-dimensional supergravity is to identify models that can be lifted to full string theory and thus enjoy an ultra-violet (UV) completion within a theory of quantum gravity.
In situations where supersymmetry is preserved, this procedure may in principle be more controlled, thanks to the properties of supersymmetric vacua per se.
On the contrary, when supersymmetry is broken the task of identifying the origin of a given four-dimensional supergravity model from string theory becomes truly challenging.
Of particular interest are de Sitter and anti-de Sitter maximally symmetric backgrounds, due to their relevance in cosmology and in holography respectively.

An intermediate step in understanding which gravitational effective field theories (EFTs) may or may not arise from string theory is the so-called swampland program (for a recent review see for example \cite{Palti:2019pca}).
It comprises a set of conjectured constraints that have to be satisfied by an EFT, if this has a chance to find a UV completion within quantum gravity.
However, the level of rigor of such constraints is varying.
Some of these conjectures have been tested on a large number of diverse examples and have been also derived in simple setups from a top-down approach, whereas others remain more speculative.
Historically, one of the earliest such conjectures is the so-called weak gravity conjecture (WGC) proposed in \cite{ArkaniHamed:2006dz}, which restricts the gauge couplings and the charges of abelian groups.
Here, of particular interest to us is its magnetic version, that dictates the existence of an upper bound on the UV cut-off of a certain EFT given by 
\begin{equation}
\label{mWGC-intro} 
\Lambda_{\rm UV} \lesssim g M_P \, , 
\end{equation}
where $g$ is the gauge coupling of an abelian gauge vector (assuming integer particle charges) and $M_P$ is the reduced Planck mass.

In this work, we consider four-dimensional N=2 supergravity models.
Indeed, gauged N=2 supergravity already provides a very rich playground in which to test various swampland conjectures.
In particular, the vacuum structure of an N=2 supergravity theory is controlled by the type of gauging one performs.
This makes it not only ideal for the study of the WGC, but also provides an explicit arena where to explore possible connections to other conjectures.
N=2 supergravity is also known for its direct relation to string compactifications on Calabi--Yau manifolds, thus further motivating its use to test conjectures on properties of quantum gravity.
On top of that, one way to get N=1 vacua from string theory is as truncations (typically orientifolds) of N=2 constructions.
In this sense, four-dimensional N=2 supergravity is also a very useful tool for strengthening, or challenging, the web of swampland conjectures.
For all these reasons, here we focus explicitly on de Sitter critical points that are obtained from N=2 supergravity coupled to vector multiplets.

A striking feature of N=2 gauged supergravity coupled to vector multiplets is that it leads to de Sitter critical points with the property 
\begin{equation}
\label{H-intro}
H \, \sim \, g \, q_{3/2} \, M_P \, \sim \, q_{3/2} \, \Lambda_{\rm UV} \, , 
\end{equation}
where $H$ is the Hubble scale, $g$ is the gauge coupling of an abelian gauge vector, $q_{3/2}$ is the abelian gravitino charge (which is quantized) and $\Lambda_{\rm UV}$ is the cut-off dictated by the magnetic WGC \eqref{mWGC-intro}.
As a result, we see that the Hubble scale is quantized in terms of the UV cut-off, indicating a dangerous sensitivity to UV corrections (either quantum or stringy) and thus making these vacua untrustworthy.
Interestingly, one can interpret such a behaviour of N=2 supergravity as a manifestation of an underlying Dine--Seiberg problem \cite{Dine:1985he}, which tells us that de Sitter vacua live near regions that get strong quantum corrections.
In addition, we will see that \eqref{H-intro} is closely related to the fact that all de Sitter vacua we consider have the specific property 
\begin{equation} 
\label{m32-intro} 
\det\,  \large(m_{3/2}\large)_{AB}  = 0 \,, 
\end{equation}
where $\large(m_{3/2}\large)_{AB}$ is the Lagrangian mass matrix of the gravitini.
This leads us to conjecture that all de Sitter critical points of gauged supergravity satisfying \eqref{m32-intro} are in the swampland.

Recently, stringent constraints on EFTs in de Sitter were put forward in \cite{Montero:2019ekk}, where the bound $m^2 \gtrsim q\, g\, H\, M_P $ is proposed for every particle in the spectrum, $m$ being the Lagrangian mass and $q$ the charge (see also \cite{Antoniadis:2020xso}).
We will not investigate this bound, which is stronger than the WGC, but one can see that it is compatible with our analysis.
Indeed, in extended supergravity a scalar potential can exist only in the presence of a gauging, which in turn means the gravitini are charged.
The fact that de Sitter vacua in gauged supergravity with vanishing gravitini masses are generically expected to violate the bound of \cite{Montero:2019ekk} can be taken as further support for our conjecture.

In the remainder of the article we will proceed by discussing in detail explicit examples, in order to give the required evidence to support the above claims.

\section{Constraints on de Sitter from WGC} 
\label{sec:constraints_on_de_sitter_from_wgc}

Before delving into the examples that give support to our analysis and conjecture, we review some aspects of the WGC, discussing its relevance in restricting de Sitter vacua in general and more specifically in gauged supergravity.

We consider a four-dimensional gravitational theory that contains an abelian vector $A_m$ with canonical kinetic term 
\begin{equation}
\label{EFT-REM}
e^{-1} {\cal L} = \frac12 M_P^2 R - \frac1{4 g^2} F_{mn} F^{mn} + \dots \, , 
\end{equation}
where $F_{mn} = \partial_m A_n - \partial_n A_m$.
Notice that the identification of $g$ as gauge coupling can be ambiguous, unless we postulate that in our EFT we deal with fields $\chi$ that are coupled to $A_m$ via the usual minimal coupling
\begin{equation}
\label{ChQuant}
\partial_m \chi + i q A_m \chi \ , \qquad q \in \mathbb{Z} \,, 
\end{equation}
where $\chi$ refers to either fermions or bosons.
Notice also that we are assuming that the charge $q$ is quantized.
Strictly speaking, this is an external input to our description, nevertheless we impose it since it is a property of the full quantum regime.

Within this setup, two conjectures have been proposed in the seminal work  \cite{ArkaniHamed:2006dz}: one is the so-called electric WGC that restricts the charges and the masses of the fields, while the other is the so-called magnetic WGC that constrains the EFT cut-off.
Here, we will mostly focus on the latter.\footnote{See also \cite{Huang:2006hc} for an explicit discussion on the magnetic WGC for de Sitter and anti-de Sitter spaces.} 
More specifically, the magnetic WGC dictates that the UV cut-off of the EFT described by \eqref{EFT-REM} is bounded by 
\begin{equation} 
\label{mWGC}
\Lambda_{\rm UV} \lesssim g M_P \,.
\end{equation}
Now, let us assume that a positive cosmological constant term is understood in the dots in \eqref{EFT-REM}.
As it will be convenient in our examples later on, we can parametrize this term by the Hubble scale $H$ as follows 
\begin{equation}
\label{HCC}
e^{-1} {\cal L}_{CC} = - 3 H^2 M_P^2 \,.
\end{equation}
Then, in general one should expect the following condition to hold
\begin{equation}
\label{HsmL}
H \ll \Lambda_{\rm UV} \, , 
\end{equation}
in order to be able to claim a good handle on the possible quantum and higher order corrections.

Since the condition \eqref{HsmL} is crucial for our work, we would like to support it with some physical arguments.
The first of these arguments is quite practical and can be illustrated by simply considering a light scalar $\phi$ with canonical kinetic terms, in the presence of a positive cosmological constant \eqref{HCC}.
It is known that quantum fluctuations of a scalar in a de Sitter background are of order $\Delta \phi \sim H/(2\pi)$.
They are essentially due to the thermal fluctuations related to the Gibbons--Hawking temperature of de Sitter (see e.g.~\cite{Carroll:2004st}).
Then, one clearly needs to have $\Lambda_{\rm UV} \gg \Delta \phi$ otherwise the quantum fluctuations would be too large and render the EFT unreliable, as all higher order operators that are suppressed by $\Lambda_{\rm UV}$ (and thus ignored in the EFT) would be in fact active and influence the observations.
Actually, for inflation one typically has $H / \Lambda_{\rm UV} \sim 10^{-5}$, as the self-consistency of the EFT description would require.

A second argument in support of \eqref{HsmL} comes from standard EFT considerations.
Since from the EFT perspective all higher order corrections are typically suppressed by the same UV cut-off, we could have higher order corrections in the gravitational sector of the form 
\begin{equation}
\label{R2}
e^{-1}{\cal L}_{grav.} =  M_P^2 \left(\frac12 R + \frac{\alpha }{\Lambda_{\rm UV}^2} R^2 + \cdots \right) \,, 
\end{equation}
where $\alpha$ is an order one parameter.
On a spatially flat de Sitter background one has 
\begin{equation}
R = 12 H^2 \, , 
\end{equation}
therefore, if we want to safely ignore the $R^2$ corrections in \eqref{R2}, we only need to ask that \eqref{HsmL} holds.
Indeed, in such a case we have $|R| \gg R^2 / \Lambda_{\rm UV}^2$ and so we can consistently keep only the two-derivative gravity sector.

Our third argument in favour of \eqref{HsmL} is more heuristic, but also more general.
We recall that an EFT is only suitable for distances that are at least greater than the corresponding de Broglie wavelength of its momentum cut-off, $\lambda_{\rm UV} \sim 1/\Lambda_{\rm UV}$.
If we are using such an EFT to probe distances comparable to $\lambda_{\rm UV}$, then we have to include also a series of irrelevant operators in the action, as long as they are allowed by the symmetries.
These operators will generically be suppressed by the UV cut-off scale $\Lambda_{\rm UV}$.
On the contrary, if we want to safely ignore the effect of all such higher dimensional operators, then we can only study distances $\lambda_{EFT}$ much larger than $\lambda_{\rm UV}$, namely
\begin{equation} 
\label{LengthsUV}
\lambda_{EFT} \gg \lambda_{\rm UV} \sim \frac{1}{\Lambda_{\rm UV}} \,.
\end{equation} 
In particular, with $\lambda_{EFT}$ we refer here to any wavelength or distance we want to describe or probe within our two-derivative supergravity EFT.
Given this, we notice rightaway that on a de Sitter background the only meaningful distances that we can probe are bounded by the de Sitter horizon radius, which means we can only have 
\begin{equation} 
\label{LengthsIR}
\lambda_{EFT} \lesssim H^{-1} \,.
\end{equation} 
Combining \eqref{LengthsUV} and \eqref{LengthsIR}, we see that our EFT is only suitable for energy scales with characteristic de Broglie wavelength within the range 
\begin{equation}
\label{LengthsEFT}
\frac{1}{\Lambda_{\rm UV}} \ll \lambda_{EFT} \lesssim \frac{1}{H} \,.
\end{equation} 
In this sense, one could figuratively say that $H$ serves as an IR cut-off for EFTs in de Sitter space.
Clearly, an EFT coupled to gravity with $H \gtrsim \Lambda_{\rm UV}$ would violate \eqref{LengthsEFT} and lead to a situation where the longest distance, which are typically the ones we believe to be protected from UV quantum corrections, would be in fact comparable to the wavelength of the UV cut-off and thus would receive non-negligible quantum corrections.

In view of these arguments, we conclude that the condition \eqref{HsmL} is a consistency condition that protects the two-derivative gravity EFT from uncontrolled UV quantum corrections.
We stress nevertheless that there is nothing wrong with probing distances that approach the wavelength corresponding to the UV cut-off, as long as one takes into account the appropriate higher dimensional operators.
However, if one wants to remain safely within the regime of validity of the two-derivative gravity action, then \eqref{LengthsUV} has to hold, in order that ignoring higher order corrections is justified.

To close this section, let us turn to the de Sitter vacua of gauged supergravity and give a schematic explanation of why they are constrained by the magnetic WGC.
As we will see in explicit examples, because of the gauging, de Sitter critical points generically have the property that 
\begin{equation}
\label{Hgq}
H \, \sim \, g \, q_{3/2} \, M_P \,.
\end{equation}
Even though we do not have a general proof of \eqref{Hgq}, we will see that it holds for all the stable de Sitter critical points that can be constructed by the use of vector multiplets only and also for all the unstable de Sitter critical points that we tested. Then, inserting \eqref{Hgq} and \eqref{mWGC} into \eqref{HsmL} the issue is manifest because it would imply
\begin{equation}
H \ll \Lambda_{\rm UV}  \ \ \to \ \  \, q_{3/2}  \ll 1 \ \,, 
\end{equation}
which is in contradiction with charge quantization \eqref{ChQuant}.
Therefore, we conclude that, since $q_{3/2}$ is quantized and cannot be arbitrary small, \eqref{HsmL} is violated and de Sitter critical points are very sensitive to higher order corrections.\footnote{With the same argument, one can show that pure Fayet--Iliopoulos terms in N=1 supergravity are in contrast with the WGC \cite{Cribiori:2020wch}.}
The rest of the article is devoted to establish that \eqref{Hgq} holds in our examples and that it is expected to hold generically, when at least one gravitino has vanishing Lagrangian mass.


\section{Abelian gaugings} 
\label{sec:abelian_gaugings}

The first class of models we consider are N=2 gauged supergravities where the gravity multiplet is coupled to an arbitrary number of vector multiplets and U(1)$_R$ Fayet--Iliopoulos (FI) coefficients are turned on.
We especially focus on cases where the scalar manifold is homogeneous, so that we can use its coset structure to derive general results.

We present our results in a series of steps: 
First we prove that de Sitter vacua with a vanishing gravitino mass matrix clash with the WGC, then we give a simple expression for the scalar potential and we identify the property that signals the vanishing of the gravitino mass matrix and finally we show how this applies to two fairly general classes of models with such properties.
From now on we set $M_P =1$.

Details on the special K\"ahler structure of the scalar manifold, its definition in terms of projective and normal coordinates, as well as other conventions we use in the following can be found in the appendix \ref{sec:gauged_n_2_supergravity}.

When considering U(1)$_R$ FI terms, the scalar potential can be expressed entirely in terms of a superpotential $W$ in a way that is reminiscent of N=1 models:
\begin{equation}\label{potential}
	{\cal V} = e^{K}\left(g^{I \bar J} D_I W \overline{D}_{\bar J} {\overline W} - 3 W \overline{W}\right),
\end{equation}
with
\begin{equation}
	D_I W = W_I + K_I W
\end{equation}
and lower indices attached to $W$ and $K$ denote partial differentiation of the superpotential $W$ and K\"ahler potential $K$.
The superpotential is in turn defined in terms of the holomorphic symplectic sections $Z^M$ and the FI terms $Q^M = \left(m^\Lambda,e_{\Lambda}\right)$ as
\begin{equation}
	W = \langle Z, Q \rangle = X^\Lambda e_{\Lambda} - F_{\Lambda} m^\Lambda.
\end{equation}
For the sake of generality we introduced both electric and magnetic charges, but we know that for any consistent gauging field redefinitions allow one to reduce to the case where only electric charges are turned on (i.e.~$m^\Lambda = 0$).
However, this also implies a rotation of the symplectic sections and the choice of a preferred basis, which may be incompatible with the existence of a prepotential $F(X)$, usually introduced as a way to define special K\"ahler geometry.
For this reason we prefer to maintain whenever possible duality covariance and allow for generic FI terms.

\subsection{The vacuum energy with vanishing gravitino mass} 

Our first statement applies to \emph{any} U(1)$_R$ gauging, regardless of the details of the scalar manifold: Whenever the Lagrangian gravitino mass matrix is vanishing at a de Sitter critical point the value of the cosmological constant is of the order of the maximum cutoff scale required by the magnetic WGC.

Since we are studying the vacuum structure, \emph{it is always understood that we are evaluating all expressions at the  critical point.}
Also, for this proof, we assume a symplectic frame where the magnetic charges are vanishing. 
As mentioned above, this is not a restrictive assumption.

Our starting point are the kinetic terms for the vector fields
\begin{equation}\label{LkinI}
{\cal L}_{kin.} = \frac14 \,e\, \, {\cal I}_{\Lambda \Sigma} \, F^\Lambda_{mn} F^{\Sigma mn} \,.
\end{equation}
The gauge kinetic functions ${\cal I}_{\Lambda \Sigma}$ form a matrix that depends on the values of the various moduli, so we cannot have any special restriction on it except that it has to have negative-definite eigenvalues such that the vacuum is ghost-free.
Since we are performing a U(1)$_R$ gauging, the formal condition that the gravitino mass matrix is vanishing gives 
\begin{equation}
\label{m32=0-Sec3}
(m_{3/2})_{AB} = 0 \quad \Leftrightarrow \quad  | X^\Lambda \vec{P}_\Lambda |  =0 \, .
\end{equation}
The vector $ \vec{P}_\Lambda$ is the triplet of the so-called moment maps, transforming in the adjoint of SU(2)$_R$, and $A,B$ are the SU(2)$_R$ indices in the fundamental.
Since we are dealing with a U(1)$_R$ gauging, one can always rotate the non-vanishing moment maps in a specific direction, say
\begin{equation}
e_\Lambda = P^3_\Lambda \, , 
\end{equation}
without loss of generality.
As a result, we have that vanishing gravitino masses mean vanishing superpotential $W = 0$ (because $m^\Lambda=0$).

As far as the vacuum energy is concerned, one can easily check that under the condition \eqref{m32=0-Sec3} the gauged N=2 scalar potential takes the form\footnote{We are using the special K\"ahler geometry relation $g^{I\bar{J}} \, U_I^\Lambda \overline{U}_{\bar{J}}^\Sigma = -\frac12 {\cal I}^{\Lambda \Sigma} - \overline{L}^\Lambda L^\Sigma$.} 
\begin{equation}
\label{V-Sec3}
\mathcal{V} = -\frac12 \, {\cal I}^{\Lambda \Sigma} \,e_\Lambda e_\Sigma > 0 \,,
\end{equation}
where ${\cal I}^{\Lambda \Sigma}$ is the inverse of the gauge kinetic matrix in (\ref{LkinI}).
Notice that, because of the negative-definite eigenvalues of ${\cal I}$, \eqref{V-Sec3} is positive definite.

Since we want to identify the gravitino charge, we turn to the covariant derivatives of the gravitini which include the following terms that are relevant for us 
\begin{equation}
D_m \psi_{n A} = \dots + \frac{i}{2}\, A_m^\Lambda \,e_\Lambda\, (\sigma^3)_A{}^B \psi_{n B} \,.
\end{equation} 
Now we identify the gauge vector $v_m$ that gauges the $U(1)_R$ by setting 
\begin{equation}
v_m = \Theta_\Lambda \, A_m^\Lambda \, , \qquad \Theta_\Lambda \equiv \frac{e_\Lambda}{2 q} \,, 
\end{equation}
and thus the gravitini have charge $\pm q$ with respect to this vector $v_m$:
\begin{equation}\label{gravitinicharge}
D_m \psi_{n A} = \dots + i\,  q \, v_m (\sigma^3)_A{}^B \psi_{n B} \,.
\end{equation} 
The vectors `orthogonal' to $v_m$ can also be easily identified by means of the projectors 
\begin{equation}
P^{\parallel \Lambda}{}_{\Sigma} = \frac{{\cal I}^{\Lambda \Gamma}\Theta_\Gamma  \Theta_\Sigma }{\Theta^2}
\ , \qquad 
P^{\bot\Lambda}{}_{ \Sigma} = \delta^\Lambda_{\Sigma} -  P^{||\Lambda}{}_{\Sigma} 
\,,
\end{equation} 
where $\Theta^2 =  \Theta_{\Lambda} {\cal I}^{\Lambda \Sigma} \Theta_{\Sigma} = \frac{{\cal V}}{2q^2}$.
Using such projectors we can split vector fields as
\begin{equation}
\label{AtoBv}
A^\Lambda_m = B^\Lambda_m + \frac{{\cal I}^{\Lambda \Gamma}\Theta_\Gamma}{\Theta^2} v_m \,,
\end{equation} 
where 
\begin{equation}
B^\Lambda_m = P^{\bot \Lambda}{}_{\Sigma} A^\Sigma_m 
\end{equation} 
satisfies $\Theta_\Lambda B^\Lambda_m = 0 = e_\Lambda B^\Lambda_m$.
Once we insert the expression \eqref{AtoBv} back into the kinetic terms we find 
\begin{equation}
e^{-1}{\cal L}_{kin.} = \frac14 \, {\cal I}_{\Lambda \Sigma} \, F^\Lambda_{mn}(B) F^{\Sigma mn}(B) 
+ \frac14 \, \frac{2 q^2}{{\cal V}} \, F_{mn}(v) F^{mn}(v) \,, 
\end{equation}
which means we obtain a canonical kinetic term for $v_m$ by setting
\begin{equation}
\label{V-Sec3-fin}
{\cal V} = 2 \, g^2 \, q^2 \,, 
\end{equation}
where $q$ is the norm of the gravitini charge, according to (\ref{gravitinicharge}), and $g$ is the gauge coupling for the canonically normalized $v_m$.

We conclude that when we have a $U(1)_R$ FI gauging and a vanishing Lagrangian mass for the gravitini then the vacuum energy is of the form \eqref{V-Sec3-fin}.
This vacuum energy gives rise to a Hubble scale of the form \eqref{Hgq} and as we have shown in the end of section \ref{sec:constraints_on_de_sitter_from_wgc} it means that such de Sitter critical points cannot be trusted as they receive huge quantum/stringy corrections according to the WGC.

\subsection{de Sitter vacua and vanishing gravitino mass} 
\label{sub:de_sitter_vacua_and_vanishing_gravitino_mass}

The general argument presented in the previous section is based on the knowledge that at the de Sitter critical point the gravitino mass is vanishing, which is equivalent to the vanishing of the superpotential $W$.
We now show that this is indeed what happens when one considers the conditions to have de Sitter critical points in ample classes of models.

Before entering into the details of the models, we recall the condition for a critical point of the potential (\ref{potential}), obtained for a generic special K\"ahler manifold and an arbitrary U(1)$_R$ gauging \cite{Fre:2002pd}:
\begin{equation}\label{cp}
	\partial_I {\cal V} = i\, \widehat{C}_{IJK}\, \overline{D}^J \overline W \,  \overline{D}^K \overline W - 2 \, e^{K}\, \overline{W}  \, D_I W = 0,
\end{equation}
where $\widehat{C}_{IJK}$ is a totally symmetric tensor specific to the manifold used.

\subsubsection{Minimal coupling}

The simplest special K\"ahler manifolds are the non-compact version of complex projective spaces, namely $\mathbb{CP}^{n_V,1} =$ U($n_V,1$)/[U($n_V$) $\times$ U(1)].
These spaces are described by the prepotential \cite{deWit:1984wbb}
\begin{equation}\label{minimal}
	F = \frac{i}{4} \eta_{\Lambda \Sigma}X^\Lambda X^\Sigma,
\end{equation}
where $\eta_{\Lambda \Sigma} = {\rm diag}\{-1,1,\ldots,1\}$.
Using normal coordinates this leads to the K\"ahler potential
\begin{equation}\label{minimalkahler}
	K = - \log \left(1 - \delta_{IJ} z^I \bar{z}^J\right)
\end{equation}
and, for the U(1)$_R$ gauging, to the superpotential 
\begin{equation}
	W = e_0 + e_I z^I +\frac{i}{2} \left(m^0 - m^I \delta_{IJ} z^J\right).
\end{equation}
The structure of the prepotential implies that the gauge kinetic functions are constant and hence the name ``minimal coupling''.
For this class of manifolds we have $\widehat{C}_{IJK} = 0$ and therefore the critical point condition becomes
\begin{equation}
	\partial_I {\cal V} = 0 \quad  \Leftrightarrow \quad (D_I W )\overline{W} = 0.
\end{equation}
This implies that either $D_IW = 0$ and therefore we have supersymmetric anti-de Sitter critical points, or $W = 0$ and we have de Sitter critical points (the case $D_IW=0=W$ leads to supersymmetric Minkowski critical points).
We therefore satisfy the general condition that de Sitter critical points need $W = 0$ and then the proof of the previous subsection applies, so that these critical points are excluded as consistent critical points of an effective theory by the WGC.

\subsection{Very special K\"ahler manifolds} 
\label{sub:very_special_k_ahler_manifolds}

The second class of theories we want to study are the symmetric very special manifolds.
These are characterized by a cubic prepotential 
\begin{equation}\label{cubicprep}
F = \frac16 C_{IJK} \frac{X^I X^J X^K}{X^0} \,,
\end{equation} 
where $C_{IJK}$ are constants related to $\widehat{C}_{IJK} = e^K C_{IJK}$.
This class contains both homogeneous and non-homogeneous manifolds.
In any case, all their isometries can be obtained by algebraic means and a full classification follows from their analysis as presented in \cite{deWit:1991nm,deWit:1992wf}, which we will closely follow in our work.

We especially decided to focus on the largest class of homogeneous very special manifolds, namely
\begin{equation}
	L(0,p) = \frac{{\rm SU}(1,1)}{{\rm U}(1)} \times \frac{{\rm SO}(2+p,2)}{{\rm SO}(2+p) \times SO(2)}, 
\end{equation}
where clearly the number of vector multiplets satisfies $n_V = 3+p \geq 3$.
Their geometry is then fixed by (\ref{cubicprep}), where the only non-vanishing entries of the $C_{IJK}$ constants are 
\begin{equation}\label{Cdef}
C_{122} = 1 \, , \quad C_{133} = - 1 \, , \quad C_{2ij} = - \delta_{ij} \, , \quad C_{3ij} = \delta_{ij}   \,,
\end{equation}
where we split $I = \{1,2,3,i\}$, with $i = 4, \ldots, 3+p$, so that
\begin{equation}
	F = \frac12 \frac{X^1}{X^0}\left[(X^2)^2-(X^3)^2\right]+ \frac12 \frac{X^3-X^2}{X^0}\,(X^i)^2.
\end{equation}
The K\"ahler potential is then given by 
\begin{equation}
K = - \log \left( \frac16 C_{IJK} \, [i (z^I - \overline z^{\overline I})]  \, [i (z^J - \overline z^{\overline J})]  \, [i (z^K - \overline z^{\overline K})] \right) \, , 
\end{equation}
and the superpotential for the most general abelian FI-gauging in normal coordinates is
\begin{equation}
W = e_0 + e_I z^I + m^0 \,\frac16 C_{IJK}z^I z^J z^K - \frac12 C_{IJK} m^I z^J z^K.
\end{equation}
Since we are not constraining the charges in the superpotential, we can use the homogeneity of the manifold to scan for the critical points at the ``origin'' of the scalar manifold 
\begin{equation}
z^1_* = - \frac{i}{2} \ , \quad z^2_* = - \frac{i}{\sqrt 2} \ , \quad z^3_* = z^i_* = 0 \,, 
\end{equation}
where 
\begin{equation}
K|_* = 0 \ , \quad K_{1}|_* = i \ , \quad K_{2}|_* = \sqrt 2 i \ , \quad K_{3}|_* = K_{i}|_* = 0 \, , 
\end{equation}
and for the metric we have the non-vanishing entries 
\begin{equation}
g_{1 \overline 1}|_* = g_{2 \overline 2}|_* = g_{3 \overline 3}|_* = 1 \ , \quad g_{i \overline j}|_* = \sqrt 2\, \delta_{ij} \,.
\end{equation}
Any critical point that for given charges does not appear at this point can always be brought back to the origin by a field redefinition which follows from the symplectic transformation acting on both the charges and the sections \cite{DallAgata:2011aa}.
In the rest of this section, since we are only interested in quantities evaluated at this point we will not clutter our formulas with the `$*$' any more, but it will be understood and we will write it only in some cases where we want to stress it.

We now prove that unless $W|_* = 0$ we cannot have a de Sitter critical point.
Let us start by assuming that 
\begin{equation}
D_I W \ne 0 \ , \quad W  \ne 0 \, .
\end{equation}
The critical point conditions (\ref{cp}) imply
\begin{equation}
\label{EOM-CIJK}
i C _{IJK} \overline D^{J} \overline W \, \overline D^{K} \overline W = 2 \overline W  D_I W \,.
\end{equation}
These conditions can now be used to derive some stronger results, by using the symmetry properties of the scalar manifold.
As shown in \cite{deWit:1991nm,deWit:1992wf}, for cubic prepotentials one always find a matrix $\tilde B$ such that 
\begin{equation}
\sum_I \tilde B^I{}_{(L}  C_{JK)I} = 0 \,.
\end{equation}
For the homogeneous very special K\"ahler manifolds defined by (\ref{Cdef}) this condition is satisfied by
\begin{equation}\label{Btilde}
\tilde B = {\rm diag}\left\{ 1 , -\frac12, -\frac12 , \frac14, \ldots, \frac14 \right\}\, , 
\end{equation}
We can therefore use such matrix, by multiplying the critical point condition with $ \overline D^{L} \overline W \tilde B^I{}_L$ such that
\begin{equation}
i\, \overline D^{L} \overline W\, \tilde B^I{}_L C _{IJK} \, \overline D^{J} \overline W \, \overline D^{K} \overline W 
= 2 \overline W \overline D^{L} \overline W \tilde B^I{}_L  D_I W \,, 
\end{equation}
which implies
\begin{equation}
\overline D^{L} \overline W\, \tilde B^I{}_L\,  D_I W = 0 \,.
\end{equation}
Using the specific expression for the $\tilde B$ matrix given in (\ref{Btilde}) we then get 
\begin{equation}
\label{FromTildeB}
|D_i W|^2 = -4 |D_1 W|^2 + 2 |D_2 W|^2 + 2 |D_3 W|^2 \,,
\end{equation}
which we can use to simplify the conditions coming from (\ref{cp}).
This can be seen by extracting from \eqref{EOM-CIJK} the components with $I=i$ 
\begin{equation}
\frac{i}{\sqrt 2} \overline D_{\overline \imath} \overline W \left( \overline D_{\overline 3} \overline W  - \overline D_{\overline 2} \overline W\right) =  
\overline W  D_i W \,, 
\end{equation}
and its complex conjugate
\begin{equation}
-\frac{i}{\sqrt 2} D_{j} W \left( D_{3} W  - D_{2} W \right) =  W  \overline D_{\overline \jmath} \overline W \,.
\end{equation}
Summing the two relations we get 
\begin{equation}
|D_i W|^2 |D_3 W - D_2 W|^2 = 2 |D_i W|^2 |W|^2 \,.
\end{equation}
If we assume $D_i W |_* \ne0$ then
\begin{equation}
|D_3 W - D_2 W|^2 = 2 |W|^2 \,.
\end{equation}
In a similar way, from the $I=1$ component of \eqref{EOM-CIJK}
\begin{equation}
i (\overline D_{\overline 2} \overline W)^2 - i (\overline D_{\overline 3} \overline W)^2 = 2 \overline W D_1 W 
\end{equation}
and its conjugate, we obtain
\begin{equation}
|D_2 W - D_3 W|^2 |D_2 W + D_3 W|^2 = 4 |D_1 W|^2 |W|^2 \, ,
\end{equation}
which, using that $|D_3 W - D_2 W|^2 = 2 |W|^2 \ne 0$, gives
\begin{equation}
|D_2 W + D_3 W|^2 = 2 |D_1 W|^2 \,.
\end{equation}
We therefore see that at the critical point, the potential can be simplified by using the condition we just derived together with \eqref{FromTildeB}, so that 
\begin{equation}
{\cal V}|_* = 3 |D_2 W|^2 + 3 |D_3 W|^2 
- \frac32 |D_2 W + D_3 W|^2 
- \frac32 |D_2 W - D_3 W|^2 
\equiv 0  \,.
\end{equation}
We conclude that if we ask that $W|_* \ne 0$ and that $D_i W|_*\ne 0$ then we can only get Minkowski critical points ${\cal V}|_* \equiv 0$.

Relaxing the condition $D_i W|_*\ne 0$  does not change the conclusion.
In fact, imposing $D_i W|_* = 0$, implies that from \eqref{FromTildeB} we now find 
\begin{equation}
2 |D_1 W|^2 =   |D_2 W|^2 +  |D_3 W|^2 \, , 
\end{equation}
which brings the vacuum energy to the form 
\begin{equation}
\label{VDi0} 
{\cal V}|_* = 3 |D_1 W|^2 - 3 |W|^2 \,
\end{equation}
and using the square of the relations following from $I=2$ in \eqref{EOM-CIJK} 
\begin{equation}
\label{DW2}
|D_2 W|^2 |D_1 W|^2 = |D_2 W|^2 |W|^2 \,
\end{equation}
and from $I=3$ 
\begin{equation}
\label{DW3}
|D_3 W|^2 |D_1 W|^2 = |D_3 W|^2 |W|^2 \,,
\end{equation}
we see that either $D_2 W \ne 0$ or $D_3 W \ne 0$, but then from \eqref{DW2} or \eqref{DW3} we have 
\begin{equation}
|D_1 W|^2 = |W|^2 \, 
\end{equation}
and finally once more ${\cal V}|_* = 0$, or $D_3 W = 0$ and $D_2 W = 0$, which further implies $|D_1 W|=0$ and finally 
\begin{equation}
{\cal V}|_* = - 3 |W|^2 \,.
\end{equation}
So, we see that if we ask that $W|_* \ne 0$ and that $D_i W|_* = 0$ then we get ${\cal V}|_* \equiv 0$ or ${\cal V}|_* = - 3 |W|^2$.

We therefore conclude that we can only find a positive vacuum energy when the gravitino masses and the superpotential vanish at the critical point $W|_* = 0$.
As we have shown in the previous subsections such de Sitter vacua of N=2 supergravity will receive huge quantum/stringy corrections.



\section{Non-abelian gaugings} 
\label{sec:non_abelian_gaugings}

It is by now a well established fact that finding (meta)stable de Sitter critical points in extended supergravity theories is extremely difficult and that even unstable critical points with small tachyon masses relative to the cosmological constant value are extremely scarce \cite{Fre:2002pd,Ogetbil:2008tk,Roest:2009tt,Borghese:2011en,DallAgata:2012plb,Catino:2013syn}.
While this could be a sign that the de Sitter conjecture \cite{Obied:2018sgi} may be somehow incorporated automatically in (extended) supergravities, it is definitely interesting to better understand the necessary and/or sufficient conditions to obtain (meta)stable vacua.
At the same time, as we want to show in this work, these same vacua may be challenged by the WGC and in turn this would leave the landscape of de Sitter vacua in extended supergravity completely empty.

After the first example of stable de Sitter vacuum was found in N=2 supergravity \cite{Fre:2002pd}, three elements were deemed necessary: non-compact gaugings, de Roo--Wagemans symplectic angles, Fayet--Iliopoulos terms.
As we will show in the following, only the third ingredient is strictly necessary, when considering arbitrary gaugings.
In fact, the de Roo-Wagemans angles are an artefact of insisting on having a purely electric gauging in the chosen symplectic frame and we can find stable vacua also without non-compact gaugings, though this type of gaugings clearly help in lifting the value of the cosmological constant due to the necessary spontaneous breaking of the gauge group to a compact subgroup on the vacuum.

If we look at the examples provided so far, a physical intuition on the construction of these vacua can be gained in analogy with the way many meta-stable de Sitter vacua have been constructed in minimal supergravity.
First one generates a runaway positive potential ${\cal V}_F$ by means of some N=2 Fayet--Iliopoulos terms and then one lifts the runaway directions with a genuine gauging term ${\cal V}_D$.
Clearly if ${\cal V}_D$ is the result of a non-compact gauging, the gauge group breaking at the vacuum generically gives a further positive contribution to the vacuum energy.
However, we will explicitly show with a new example that one can also obtain metastable critical points with a pure compact gauging resulting entirely from N=2 Fayet--Iliopoulos terms.

Once again, in this work we concentrate ourselves on N=2 supergravity coupled to vector multiplets, leaving a discussion on more general couplings and/or more supersymmetries for the final section.
We will see that in all the examples we present the WGC signals their incompatibility with quantum gravity.

Before entering into the details of the examples, let us recall a few important facts about non-abelian gaugings (more details can be found in the appendix \ref{sec:gauged_n_2_supergravity}).
The scalar potential is constructed by the sum of two terms, one coming from the triplet of Fayet--Iliopoulos charges $Q^{M x} = \left(P^{\Lambda x}, P_{\Lambda}^x\right)$, with $x=1,2,3$,
\begin{equation}\label{Fterm}
	{\cal V}_F = g^{I \bar J} \langle U_I ,Q^x\rangle \langle  \overline{U}_{\bar J}, Q^x\rangle - 3 \langle V, Q^x\rangle \langle \overline V, Q^x\rangle,
\end{equation}
and one from the proper gauging of the isometries of the scalar manifold, generated by the holomorphic Killing vectors $k_M^I$:
\begin{equation}\label{Dterm}
	{\cal V}_D = |\overline{V}^M k_M|^2 = \overline{V}^M k_M^I V^N \bar{k}_N^{\bar{J}} \,g_{I \bar J}.
\end{equation}
When the FI charges are turned on in a single U(1) direction within the SU(2)$_R$ (namely when $Q^{Mx}$ can be rotated such that they are non-zero only for a single value of $x$), then ${\cal V}_F$ coincides with the potential (\ref{potential}).
We remind that the isometries have a linear action on the holomorphic sections and therefore one can also rewrite the D-term potential (\ref{Dterm}) in terms of prepotentials
\begin{equation}
	P_M^0 = e^K\, T_{MN}{}^Q \Omega_{QP} Z^N \overline{Z}^P,
\end{equation}
satisfying $k_M^I =i\, g^{I \bar J} \overline{\partial}_{\bar J} P_M$, where $T_M$ are the matrix representations of the isometry action.

\subsection{Minimal couplings} 
\label{sub:minimal_couplings}

The first example we present is again in the theory with minimal couplings, where the prepotential is defined as in (\ref{minimal}), and the (unstable) de Sitter vacua had been first uncovered in \cite{deWit:1984wbb}.
The model in \cite{deWit:1984wbb} uses the $\mathbb{CP}^{3,1}$ scalar manifold and gauges a SU(2) group diagonal between SU(2)$_R$ (gauged by the FI terms) and the SU(2) $\simeq$ SO(3) isometry group that rotates the normal coordinates $z^I$, $I=1,2,3$.

In this case we have a purely electric gauging. 
In normal coordinates the covariantly holomorphic sections are $L^0 = e^{K/2}$, $L^I = e^{K/2} z^I$, where $K = - \log\left(1 - \delta_{IJ}z^I \bar{z}^{\bar{J}}\right)$ as in (\ref{minimalkahler}).
It is however simpler to express all relevant quantities in terms of the covariantly holomorphic sections (where we raise and lower indices using the Kronecker delta). 
Moreover, since only the $L^I$ sections enter in the following expressions, we will often use the vector notation $\vec{L} = \{L^1, L^2, L^3\}$.
In details, we note the useful relation
\begin{equation}
	e^K = 1 + |\vec L|^2,
\end{equation}
which allows us to write the metric and inverse as
\begin{equation}
	g_{I \bar{J}} = e^{K}\left(\delta_{IJ} + \bar{L}_I L_J\right), \qquad g^{I \bar{J}} = e^{-K}\left(\delta^{IJ} - e^{-K}L^I \bar{L}^J\right).
\end{equation}
We also notice that
\begin{equation}
	U_I{}^J = D_I L^J = e^{K/2}\left(\delta_I^J + \bar{L}_I L^J\right),
\end{equation}
which is going to be useful to simplify the expression of the scalar potential.
The FI-term contribution is directly computed from (\ref{Fterm}) by setting $P_I^x = q \, \delta_I^x$, with any other charge set to zero and from (\ref{Dterm}), by gauging the isometries
\begin{equation}
	\delta z^I = q\, \epsilon^{IJK} \Lambda^J z^K., 
\end{equation}
for real parameters $\Lambda^I$, so that the Killing vectors are
\begin{equation}
	k_I^J = q\, \epsilon^{IJK} z^K,
\end{equation}
with corresponding prepotentials
\begin{equation}
	P_I^0 = -i\, q \, \epsilon_{IJK} \bar{L}^J L^K.
\end{equation}
Combining the resulting terms and using the relations above, the scalar potential reduces to
\begin{equation}
	{\cal V} = q^2\left(3 - 2 |\vec{L}|^2 + 2 |\vec{L}|^4 - \vec{L}^2 \overline{\vec{L}}^2\right).
\end{equation}

The critical point condition is easily computed and gives
\begin{equation}
	\partial_I {\cal V} = q^2 \,e^{K/2}\left[\overline{L}_I(2 |\vec L|^4 -2 - \vec{L}^2 \overline{\vec{L}}^2) - L_I \overline{\vec{L}}^2 \right] = 0,
\end{equation}
which shows two critical points, one where the SU(2) is preserved at
\begin{equation}
	L^I = 0, \qquad {\rm unbroken\ SU(2)}, \qquad {\cal V} = 3 \,q^2,
\end{equation}
and one where the SU(2) group is broken at
\begin{equation}
	\vec{L}^2 = 0, \quad |\vec{L}|^2 = 1, \qquad {\rm broken\ SU(2)}, \qquad {\cal V} = 2\, q^2.
\end{equation}
Both critical points are unstable as they show some negative eigenvalue in the scalar mass matrix \cite{deWit:1984wbb}.

It is also interesting to notice that the gravitino mass matrix in both critical points has at least one vanishing eigenvalue.
In details, the gravitino mass matrix is
\begin{equation}
	(m_{3/2})_{A}{}^B = \frac{i}{2}\, \langle V, Q^x\rangle (\sigma^x)_A{}^B
\end{equation}
and in our example it reduces to
\begin{equation}
	(m_{3/2})_{A}{}^B = \frac{i}{2}\, q\, L^I (\sigma^I)_A{}^B,
\end{equation}
giving
\begin{equation}
	{\rm{det}}(m_{3/2}) = \frac{q^2}{4}\, \vec{L}^2.
\end{equation}
This implies that in the most symmetric vacuum it vanishes because of the vanishing of the sections and in the other vacuum it vanishes because of the vacuum condition $\vec{L}^2 = 0$, though one of the masses may be non-vanishing.

In order to check that our conjecture works, we need to compare the charge $q$ appearing in the potential with the effective charge appearing in the covariant derivative of the gravitinos.
The general coupling of the gravitino to the gauge vectors is 
\begin{equation}
	D_m \psi_{n A} = \ldots + \frac{i}{2} \, \left[A^I_m P_I{}^x (\sigma^x)_A{}^B +  A^I_m\, P_I^0\, \delta_A^B\right] \psi_{nB},
\end{equation}
where the first contribution in the bracket comes from the FI-terms and the second from the SU(2) gauging.
For our example this becomes
\begin{equation}
	D_m \psi_{n A} = \ldots + \frac{i}{2} \, q\,\left[A^I_m (\sigma^I)_A{}^B -i \, A^I_m \epsilon_{IJK} \bar{L}^J L^K\, \delta_A^B\right] \psi_{nB}.
\end{equation}
On the SU(2) symmetric vacuum this is trivially reduced to the standard SU(2) coupling for the vectors $A^I$, with charge $q$.
For the vacuum with broken symmetry we see that once we split $\vec{L} = \vec{R} + i \, \vec{I}$, the vacuum condition implies $|\vec{R}| = |\vec{I}| = \frac{1}{\sqrt2}$ and $\vec{R}\cdot\vec{I} = 0$. 
We can then rewrite the gravitino coupling as
\begin{equation}
	D_m \psi_{n A} = \ldots + \frac{i}{2} \, q\,\left[\vec{A}_m \cdot \vec{\sigma}_A{}^B -i \, \vec{A}_m \cdot (2i\vec{R}\wedge\vec{I}) \delta_A^B\right] \psi_{nB} =  \ldots + \frac{i}{2} \, q\,\left[\vec{A}_m \cdot (\vec{\sigma})_A{}^B + A^\perp_m  \delta_A^B\right] \psi_{nB},
\end{equation}
where $A^\perp_m$ is the vector field orthogonal to the plane defined by $\vec{L}$.

We therefore conclude once more that the quantized charge should be $q$ and the vacuum is proportional to $q$ with a coefficient of order 1.


\subsection{Non-compact gaugings} 
\label{sub:non_compact_gaugings}

As we mentioned in the introduction to this section, the first example of a stable de Sitter vacuum in extended supergravity had been obtained in \cite{Fre:2002pd}, then generalized in \cite{Ogetbil:2008tk}.
We will now show that also these examples have a cosmological constant that is comparable to the WGC cutoff for a consistent effective theory that could be coupled to quantum gravity.

To be specific, we focus once more on the $L(0,p)$ manifolds described in section \ref{sub:very_special_k_ahler_manifolds}.
As explained in the introduction, a suitable choice of FI-terms gives a runaway or a no-scale, semi-positive defined scalar potential.
We then introduce a D-term to lift the (runaway) flat directions and to create a vacuum.
This requires a gauging under which all, or most of the scalar fields are charged, which, as we will see, is intimately related to the existence of non-perturbative symmetries for our examples.

In the case of special K\"ahler manifolds, and especially for the very special case, all isometries can be fully analyzed and related to the form of the prepotential, as shown by de Wit and Van Proeyen in  \cite{deWit:1991nm,deWit:1992wf}.
This immediately explains why it is useful to introduce a full symplectic-invariant formulation of N=2 supergravity and why one cannot simply reach our goal with a purely electric gauging in the standard frame.
Ordinary perturbative symmetries rotate only the $X^\Lambda$ sections and hence they act linearly on the scalar fields.
This implies that it is very difficult to stabilize all scalars unless one is allowed to introduce a very large gauge group.
However, there are also non-perturbative symmetries mixing $X^\Lambda$ and $F_{\Lambda}$ so that their gauging introduces a dependence on all scalar fields in the D-term potential.
Clearly one can once again reach the wanted result by a purely electric gauging, provided one performs a duality rotation on the symplectic base as to redefine $\widehat X^\Lambda = \widehat X^{\Lambda}\left(X^\Sigma,F_\Sigma\right)$ and go to a frame where a prepotential does not exist. 
We present all the details of the procedure in appendix \ref{sec:construction_of_the_potential}, while here we discuss the aspects relevant for our argument.

The full gauging generates a U(1)$_R \times$ SO(2,1) group. 
We go to a convenient symplectic frame where the U(1)$_R$ factor is gauged by the graviphoton and the SO(2,1) group by the first three vector fields in the vector multiplets.
To do so we introduce a symplectic rotation $S$ with respect to the basis where the cubic prepotential is defined, so that the F-term potential follows from the superpotential\footnote{In our presentation we reabsorb the overall coupling constant $g$ associated to the gauging procedure in \cite{Andrianopoli:1996cm} in the definition of the charges, prepotentials and Killing vectors.}
\begin{equation}
	W = \langle SZ, Q \rangle = e_0 (\cos \delta\, X^4 + \sin \delta\, F_4).
\end{equation}
The D-term potential on the other hand, is more complicated and all the details about the isometries and associated prepotentials are presented in the appendix \ref{sec:construction_of_the_potential}.
If we truncate to the imaginary part $y^I$ of the scalar fields $z^I$ we can produce the compact expression
\begin{equation}
	{\cal V}_D|_{\Re z^I = 0} = \frac{e^{2 K}}{4} \, e_1^2\, (y_2 - y_3)\left[(y_1+y_2+y_3)^2-(-y_1+y_2+y_3)^2(2+4y_1(y_2+y_3) - \delta_{ij}y^iy^j)\right]^2.
\end{equation}

Let us now turn to the analysis of the de Sitter critical point.
The full scalar potential indeed has a critical point at 
\begin{equation}
	z^1 = (z^2+z^3) = \frac{i}{\sqrt2}\,, \quad 	(z^2-z^3) = \cot \delta - 2 i\, \frac{e_1}{e_0}\, \csc \delta, \quad z^i = 0.
\end{equation}
At this critical point the cosmological constant value is
\begin{equation}
	\mathcal{V} = 2 \,\sin \delta\, e_0 e_1
\end{equation}
and all scalar masses are positive, but for the two goldstone bosons associated to the breaking of the gauge group
\begin{equation}
	{\rm U}(1)_R \times {\rm SO}(2,1) \to 	{\rm U}(1)_R \times {\rm U}(1).
\end{equation}
We therefore have a stable de Sitter vacuum with two residual U(1) gauge groups, which can be used to test its compatibility with the magnetic WGC.

In order to test our conjecture, we need first to redefine the vector fields and the charges to reproduce standard couplings.
The non-minimal couplings of the vector fields generate a non-trivial kinetic matrix, which however is diagonal at the critical point:
\begin{equation}
	{\cal I} = - \frac{1}{2\sin \delta} \,{\rm diag}\left\{\frac{e_0}{e_1}, \frac{e_1}{e_0}, \ldots, \frac{e_1}{e_0}\right\}.
\end{equation}
This implies that the two gauge fields for the residual U(1) factors have kinetic terms
\begin{equation}
	- \frac{1}{8\sin \delta} \frac{e_0}{e_1} \,F_{\mu \nu}^0 F^{0 \,\mu \nu} - \frac{1}{8\sin \delta} \frac{e_1}{e_0} \,F_{\mu \nu}^3 F^{3 \,\mu \nu}.
\end{equation}
Also, the triplet of prepotentials from the gauged isometries in this new basis and on the vacuum reduces to 
\begin{equation}
	P^0_{1} = P^0_2 = 0, \quad P^0_3 = e_1,
\end{equation}
which means that the gravitino is charged under the U(1)$^2$ gauge group as follows:
\begin{equation}
	D_m \psi_{n A} = \ldots + \frac{i}{2} \, \left[A^I_m P_I{}^x (\sigma^x)_A{}^B +  A^I_m\, P_I^0\, \delta_A^B\right] \psi_{nB} = \ldots + \frac{i}{2} \,\left[e_0\, A^0_m (\sigma^3)_A{}^B + e_1\, A^3_m\, \delta_A^B\right] \psi_{nB} .
\end{equation}

We can now reproduce canonical couplings by rescaling the vector fields $A^0$ and $A^3$ and introducing the charges $q_0$ and $q_1$ as follows
\begin{eqnarray}
	&&e_0 A^0 = q_0 {\cal A}^0, \qquad e_1 A^3 = q_1 {\cal A}^3, \\[2mm]
	&&\frac{1}{8\sin \delta} \frac{e_0}{e_1} \left(\frac{q_0}{e_0}\right)^2 = \frac{1}{8\sin \delta} \frac{e_1}{e_0} \left(\frac{q_1}{e_1}\right)^2= \frac14,
\end{eqnarray}
whose solution forces $q_0 = q_1 = q$.
After these rescalings we obtain canonical kinetic terms for the vectors ${\cal A}^\Lambda$ and canonical couplings to the gravitino.
We can finally look at the scalar potential expressed in terms of the ``quantized'' charge $q$ and we see that it is simply given by a product of charges (once $M_P$ has been normalized to 1)
\begin{equation}
	\mathcal{V} = q^2.
\end{equation}
We therefore conclude once again that this vacuum energy is at the threshold of the effective theory as dictated by the magnetic WGC.

For what concerns the gravitino masses, we have to check the superpotential at the critical point and we see that $W=0$, which means that both gravitinos are massless, as expected from our general discussion presented in the introduction.


\subsection{Stable vacua with SU(2)$_R$ gauging} 
\label{sub:stable_vacua_with_su_2__r_gauging}

We now display an example of a stable de Sitter vacuum resulting from the scalar potential obtained by means of a full set of SU(2)$_R$ FI terms.
The model used in this section is also a homogeneous manifold with cubic prepotential, but it does not belong to a one-parameter family, instead it is a special case.
The scalar manifold we choose is
\begin{equation}
	\frac{{\rm Sp}(6,{\mathbb R})}{{\rm U}(3)}
\end{equation}
and its prepotential is \cite{deWit:1991nm,deWit:1992wf}
\begin{equation}
	\begin{split}
	F(X) = \frac{1}{2 X^0}&\left[X^1\left((X^2)^2-(X^3)^2-(X^4)^2\right)-X^2\left((X^5)^2+(X^6)^2\right) \right. \\[2mm]
	&\left. +X^3\left((X^5)^2-(X^6)^2\right) +2 X^4X^5X^6\right]\,.
	\end{split}
\end{equation}

A quick analysis of the SU(2)$_R$ gauging of this manifold shows various families of stable de Sitter vacua.
We will not enter into the details of these calculations, but briefly outline the strategy and show the relevant details of a significant example (we checked that the other vacua we found do not give different results concerning the WGC bounds).
First of all, we employed once again the strategy of \cite{DallAgata:2011aa} to search for critical points of the potential on homogeneous manifolds.
We therefore assumed that the fixed point was at the ``origin'' of the manifold, namely at
\begin{equation}
z^1_* = - \frac{i}{2} \ , \quad z^2_* = - \frac{i}{\sqrt 2} \ , \quad z^3_* = z^i_* = 0 \,, 
\end{equation}
while leaving the FI-terms $Q^{Mx}$ arbitrary.
We then chose an electric charge along $x=1$ and a magnetic charge along $x=2$, such that the scalar potential resulting from these terms is the sum of the squares of these charges, while we left all charges available in the direction $x=3$.
As mentioned above, the critical point condition gives various families of solutions.
We focused on a specific instance where it was easy to show that there is a symplectic rotation that brings the prepotentials to the appropriate form for the SU(2)$_R$ gauging, namely
\begin{equation}\label{su2struct}
	P_{\Lambda}^x P_{\Sigma}^y \epsilon^{xyz} = q\,\epsilon_{\Lambda \Sigma \Gamma} P_{\Gamma}^z,
\end{equation}
while leaving the kinetic matrix diagonal (at the same base point).

To be specific, (in the original basis) we chose the charges
\begin{equation}
	Q_3^1 = \frac{q}{\sqrt2},\quad Q^{42} =\sqrt2 \,q,\quad Q^{23} = \frac{q}{\sqrt2}\,,\quad Q^{33} = -q\,, \quad Q_{0}^3 = \frac{q}{4\sqrt2}
\end{equation}
and the symplectic rotation
\begin{equation}
	S = \left(\begin{array}{cccccccccccc}
	 0 & \frac{1}{2\sqrt2} & \frac12 & 0 & 0 & 0 & \sqrt2 & 0 & 0 & 0 & 0 & 0\\
	 0 & 0 & 0 & 0 & 0 & 0 & 0 & 0 & 0 & \sqrt2 & 0 & 0\\
	 0 & 0 & 0 & 0 & \frac{1}{\sqrt2} & 0 & 0 & 0 & 0 & 0 & 0 & 0\\
	 0 & \frac{1}{2\sqrt2} & -\frac12 & 0 & 0 & 0 & \sqrt2 & 0 & 0 & 0 & 0 & 0\\
	 0 & 1 & 0 & 0 & 0 & 0 & -4 & 0 & 0 & 0 & 0 & 0\\
	 0 & 0 & 0 & 0 & 0 & {\mathbb 1}_2 & 0 & 0 & 0 & 0 & 0 & 0\\
	 -\frac{1}{4\sqrt2} & 0 & 0 & 0 & 0 & 0 & \frac{1}{\sqrt2} & 1 & 0 & 0 & 0 & 0\\
	 0& 0 & 0 & -\frac{1}{\sqrt2} & 0 & 0 & 0 & 0 & 0 & 0 & 0 & 0\\
	 0& 0 & 0 & 0 & 0 & 0 & 0 &0  & 0 & 0 &  \sqrt2& 0\\
	 -\frac{1}{4\sqrt2} & 0 & 0 & 0 & 0 & 0 & 0 & \frac{1}{\sqrt2} & -1 & 0 & 0 & 0\\
	 \frac18 & 0 & 0 & 0 & 0 & 0 & 0 & \frac12 & 0 & 0 & 0 & 0\\
	  & 0 & 0 & 0 & 0 & 0 & 0 & 0 & 0 & 0 & 0 & {\mathbb 1}_2
	\end{array}\right).
\end{equation}
This results in a new symplectic frame where the charges are all electric and such that
\begin{equation}
	P_{\Lambda}^x = q\, \delta_{\Lambda}^x,
\end{equation}
therefore satisfying (\ref{su2struct}).
The potential at the critical point has a positive cosmological constant
\begin{equation}
	{\cal V} = \frac32\, q^2
\end{equation}
and the scalar masses are all positive: $m^2 = 2 q^2 >0$.
The chosen symplectic frame has also the bonus of having canonically normalized kinetic terms, so that the charge $q$ corresponds precisely to the coupling of the gauge vectors to the gravitinos.
So, once more we have a scalar potential that at the de Sitter vacuum has a cosmological constant value that is at the threshold dictated by the magnetic WGC.

It is easy to check that also in this case the gravitino mass matrix vanishes at this critical point and hence once more we seem to find a relation between the vanishing of the gravitino masses and the possible violation of the magnetic WGC.



\section{Comments and outlook} 
\label{sec:comments_and_outlook}

In this work we analyzed de Sitter critical points appearing in N=2 gauged supergravity theories coupled to vector multiplets.
We have shown that in large classes of models and especially for all known stable de Sitter vacua, the value of the cosmological constant at the critical point is proportional to the square of the charge of the gauging and that therefore its scale is comparable to the cutoff scale dictated by the magnetic WGC.
We also proved that in all these examples there is at least one gravitino that is massless and therefore we argue that all de Sitter critical points with this property are in the swampland. 

When incorporated in the broader swampland picture, our results indicate that the WGC is stronger than the de Sitter conjecture \cite{Obied:2018sgi}. Indeed, we showed how the WGC can be used to exclude de Sitter critical points independently from their stability.  It is worth mentioning that the de Sitter conjecture \cite{Obied:2018sgi} has been found to be in tension with the Higgs potential \cite{Denef:2018etk} and thus refined in order to allow for local maxima, provided they are sufficiently steep \cite{Ooguri:2018wrx,Andriot:2018wzk,Garg:2018reu,Andriot:2018mav,Rudelius:2019cfh}. However, our conclusion is that in supergravity de Sitter extrema are always in the swampland as a consequence of the WGC and charge quantization alone, irrespectively of the preferred formulation of the de Sitter conjecture.

We should also stress, though, that while the supergravity models we analyzed might be in the swampland, they could still be valid consistent truncations of higher-dimensional theories.
It is a well-known fact, since \cite{DallAgata:2005zlf}, that one can have consistent truncations of flux compactifications that are not good effective theories and this may still well be the case for models having de Sitter critical points.

Natural extensions of this work move the analysis to the general couplings of N=2 supergravity also in the presence of hypermultiplets and to higher supersymmetric models.
For what concerns the hypermultiplet couplings, one should look carefully into the examples of \cite{Fre:2002pd,Ogetbil:2008tk} and especially \cite{Catino:2013syn}.
While we leave this for future work, we see that in the case of \cite{Fre:2002pd,Ogetbil:2008tk} we do not expect qualitative differences with respect to the case without hypers and indeed we can also see directly from their results that at the de Sitter vacuum the gravitino mass is vanishing.

In theories with a higher number of supersymmetries the de Sitter critical points are even scarcer than in the N=2 case.
One particularly interesting example has been proposed in \cite{DallAgata:2012plb}, where a one-parameter family of theories with SO(4,4) gauge group has been analyzed and the appearance of a de Sitter critical point with parametrically small tachyonic masses has been found.
We notice that also in this case the (in)stability of the vacuum is crucially dependent on the choice of symplectic frame and it is therefore essential that there is a one-parameter family of theories with such gauge group, as follows by applying the same technique as in \cite{DallAgata:2012mfj} for their consistent symplectic embedding.
Interestingly, the vacuum with unbroken SO(4) $\times $ SO(4) gauge group, which exists for any value of the parameter, has vanishing gravitino masses, while the vacuum with residual SO(3) $\times$ SO(3) gauge symmetry, has non-vanishing gravitino masses.
This makes it a suitable candidate to check if the condition we described on the gravitino masses is necessary and/or sufficient to obtain vacua with cosmological constant within the magnetic WGC bounds.


\section*{Acknowledgments}

\noindent 
We thank D.~Andriot, L.~Martucci, M.~Peloso and T. Van Riet for discussions.
We especially thank C.~Scrucca and G.~Zoccarato for past collaborations that led to some of the preliminary results on which sections \ref{sub:non_compact_gaugings} and \ref{sub:stable_vacua_with_su_2__r_gauging} are based.
\noindent
GD is supported in part by MIUR-PRIN contract 2017CC72MK003, FF is supported by the STARS grant SUGRA-MAX. NC is supported by an FWF grant with the number P 30265. 

\appendix 

\section{Gauged N=2 supergravity} 
\label{sec:gauged_n_2_supergravity}

In this appendix we collect some formulae about special K\"ahler geometry, which is the geometry of the vector multiplet scalar manifold in N=2 supergravities, and about the gauging procedure.
It is by no means a comprehensive summary of gauged N=2 supergravity theories, but it contains everything that is required to reproduce the calculations in this work.
For the derivation of what follows we refer the reader to the original work we used to prepare this work, namely \cite{Strominger:1990pd,Ceresole:1995jg,Andrianopoli:1996cm,Ceresole:1995ca}.

We should start by stressing that although a full N=2 duality covariant supergravity action has not been built so far, decisive steps have been taken in this direction, especially in the case of supergravity coupled to vector multiplets. 
As shown in \cite{DallAgata:2003sjo}, whenever one introduces magnetic gaugings, tensor multiplets have to be introduced.  
In the case of supergravity coupled to vector multiplets, one has therefore to improve couplings to vector-tensor multiplets.  
In \cite{DAuria:2004yjt,Andrianopoli:2007ep} the  authors  worked  out  the  supersymmetry  transformations  and  scalar  potential  for supergravity coupled to vector-tensor multiplets and for a generic gauging, although in the case of vanishing FI terms. 
The extension to non-trivial FI terms is, however, straightforward and, following \cite{DallAgata:2010ejj}, we will provide in the following the relevant quantities for our analysis.
We should also stress that an outline of the general procedure by using the embedding tensor formalism can be found in \cite{Louis:2009xd} and that general lagrangians for N=2 conformal supergravity theories with arbitrary gaugings have been presented in \cite{deWit:2011gk}.

A special K\"ahler manifold is parameterized by complex coordinates $z^I$, $I=1,\dots,n_V$.
Since this is the geometry of the vector-multiplet sector, electric-magnetic duality plays a role in constraining the manifold and this is made manifest by describing the geometry by means of holomorphic sections
\begin{equation}
Z^M = \begin{pmatrix} X^\Lambda(z) \\ F_\Lambda(z)  \end{pmatrix}  \ , \qquad \Lambda = 0, I \, ,
\end{equation} 
where the additional sections with index 0 have been added to take into account the graviphoton and its dual, which do not have corresponding scalars in their multiplet.
When a prepotential $F(X)$ exists, these sections can also be thought of as projective coordinates and $F_{\Lambda} = \partial_{\Lambda} F(X)$.
However, special geometry can be defined in the absence of such prepotential and, unless specified otherwise, we will not assume the sections are chosen in such specific frame.
Note that two different patches of the manifold are related by
\begin{equation}\label{patches}
	Z'(z) = e^{-f(z)} S Z(z),
\end{equation}
where $S$ is a constant symplectic matrix and $f$ is a holomorphic function of the coordinates, generating the K\"ahler transformations of the K\"ahler potential.
Defining the symplectic product
\begin{equation}
\langle A, B\rangle = A^T \Omega B =A^\Lambda B_\Lambda - B^\Lambda A_\Lambda,
\end{equation}
the K\"ahler potential is then
\begin{equation}
K = - \log\left[ -i \langle Z, \bar Z \rangle\right]
\end{equation}
and changing patches, from (\ref{patches}), we get the usual K\"ahler transformation
\begin{equation}\label{kaehler}
	K'(z,\bar{z}) \to K(z,\bar{z}) + f(z) + \bar{f}(\bar{z}).
\end{equation}
On the Hodge bundle over the manifold one can also define covariantly-holomorphic sections
\begin{equation}
V^M = e^\frac{K}{2}Z^M
\end{equation}
such that the whole geometric structure gets encoded in the following algebraic and differential constraints:
\begin{align}
\langle V, \bar V \rangle &= i,\\
U_I &= D_I V = (f_I^\Lambda, h_{I \Lambda}),\\
D_I U_J &= i \, \hat C_{IJK}\,g^{K\bar K}\bar U_{\bar K},\\
D_I \bar U_{\bar J} &= g_{I \bar J}\bar V,\\
D_I \bar V &=0,
\end{align}
where now $D_I$ is the covariant derivative with respect to the usual Levi--Civita connection and the K\"ahler connection $\partial_I K$. 
This means that under a K\"ahler transformation \ref{kaehler}, a generic field $\chi^I$, with charge $p$, namely transforming as $\chi^I \to e^{-\frac{p}{2}f+\frac{\bar{p}}{2}\bar{f}} \chi^I$, has covariant derivative
\begin{equation}
	D_I \chi^J = \partial_I \chi^J  + \Gamma_{JK}^I \chi^K +\frac{p}{2} \, \partial_J K \, \chi^I,
\end{equation}
and analogously for $\overline{D}_{\bar{J}}$, with $p \to \bar{p}$.
We followed standard conventions and chose $p = -\bar{p} = 1$ for the weight of $V$.
Note also that
\begin{equation}
	g_{I\bar J} = i\, \langle U_I , \overline{U}_{\bar J}\rangle.
\end{equation}

Two more ingredients needed are the matrix defining the non-minimal couplings of the vector multiplets
\begin{equation}
	{\cal N}_{\Lambda \Sigma} = {\cal R}_{\Lambda \Sigma} + i\, {\cal I}_{\Lambda \Sigma} =  \left(M_{\Lambda}, \overline{h}_{\bar{I}}\right) \left(L^\Sigma, \overline{f}_{\bar{I}}^\Sigma\right)^{-1},
\end{equation}
and the matrix
\begin{equation}
	{\cal M}_{MN} = \left(\begin{array}{cc}
	{\cal I}_{\Lambda \Sigma} + {\cal R}_{\Lambda \Delta} {\cal I}^{\Delta \Gamma} {\cal R}_{\Gamma} &  {\cal R}_{\Lambda \Delta}I^{\Delta \Sigma} \\[2mm]
	{\cal I}^{\Lambda \Gamma}{\cal R}_{\Gamma \Sigma} & {\cal I}^{\Lambda \Sigma}
	\end{array}\right),
\end{equation}
which is used in the definition of the scalar potential.
In fact, the kinetic lagrangian for the vector multiplets is
\begin{equation}
	L_{kin} = \frac14 \,e\, {\cal I}_{\Lambda \Sigma} \,F^\Lambda_{\mu\nu} F^{\Sigma \mu\nu}  + \frac18 \, {\cal R}_{\Lambda \Sigma}\, \epsilon^{\mu\nu\rho\sigma} \,F_{\mu \nu}^\Lambda F^\Sigma_{\rho \sigma},
\end{equation}
which means that ${\cal I}$ is negative definite.

The scalar potential following from the gauging procedure has two main contributions. 
The first one ${\cal V}_F$ is coming from the $N=2$ Fayet--Iliopoulos terms, which are the relics of the possible coupling to hypermultiplets.
If we consider full symplectic invariance, the FI terms are given in terms of the triplet of FI charges vectors $Q^{M x} = \left(P^{\Lambda x}, P_{\Lambda}^x\right)$, with $x=1,2,3$:
\begin{equation}\label{Ftermapp}
	{\cal V}_F = g^{I \bar J} \langle U_I , Q^x \rangle \langle  \overline{U}_{\bar J}, Q^x\rangle - 3 \langle V, Q^x\rangle \langle \overline V, Q^x\rangle.
\end{equation}

The second contribution is the $D$-term ${\cal V}_D$ generated by the proper gauging of the isometries of the scalar manifold.
Again, trying to be general and maintaining symplectic invariance, for special K\"ahler manifolds, the isometries can be derived by looking at their linear action on the sections.
In fact all isometries must preserve (\ref{patches}) and therefore
\begin{equation}
	\delta_P Z^M = (T_P)_N{}^M Z^N - f_P(z) Z^M,
\end{equation}
where $T_P$ is a symplectic matrix (the generator of $S$)
\begin{equation}\label{symplcond}
	T_{\Lambda}^T \Omega + \Omega T_{\Lambda} = 0,
\end{equation}
and $f_N(z)$ are compensating holomorphic functions, which are going to be related to how the K\"ahler potential transforms under such isometries.
Using full Sp($2n_V+2$,${\mathbb R}$) indices:
\begin{equation}\label{symplconstraint}
	T_{M[N}{}^Q \Omega_{P]Q} = 0.
\end{equation}
Consistency of the gauging also requires
\begin{equation}\label{repconstraint}
	T_{(MN}{}^Q \Omega_{P)Q} = 0.
\end{equation}
Note that now the position of the index transforming with $S$ is fixed, so that indices $M,N,\ldots$ are lowered and raised with the symplectic matrix. 
Upper indices transform with $S$ and lower indices transform with $S^{-1} = -\Omega S^T \Omega$, so that $V^M W_M = V^M \Omega_{MN} W^N$ is symplectic invariant
\begin{equation}
	V^{M\prime} W_M^\prime = V^{M\prime} \Omega_{MN} W^{N\prime} = V^P S^M{}_P \Omega_{MN} S^N{}_Q W^Q = V^M \Omega_{MN} W^N = V^M W_M.
\end{equation}
The non-linear action on the coordinates can be obtained by means of holomorphic Killing vectors, which can be related to the linear action above in frames where the prepotential exists.
In this case the Killing vectors follow by introducing \emph{normal coordinates} $z^I \equiv X^I/X^0$:
\begin{eqnarray}
	\delta_M z^I &=& \frac{\delta_{\Lambda} X^I}{X^0} - \frac{X^I}{X^0} \frac{\delta_{M} X^0}{X^0} =  \nonumber \\
	&=& \frac{(T_{M} Z)^I}{X^0} - f_{M} \frac{X^I}{X^0} - \frac{X^I}{X^0}\frac{(T_{M} Z)^0}{X^0} + f_{M} \frac{X^I}{X^0} \nonumber \\
	&=& \frac{(T_{M} Z)^I}{X^0} - \frac{X^I}{X^0}\frac{(T_{M} Z)^0}{X^0} \equiv k^I_{M}(z).
\end{eqnarray}
At the infinitesimal level
\begin{eqnarray}
	\delta_M V^N &=& - T_{MP}{}^M V^P, \\[2mm]
	\delta_M W_N &=& T_{MN}{}^P W_P.
\end{eqnarray}
Under an isometry the K\"ahler potential transforms as
\begin{equation}
	\delta_{M} K = - e^{K} \, i\,  ( \delta_{M}Z^T \Omega \overline{Z} +  Z^T \Omega \delta_{M}\overline{Z}) =  f_M + \overline{f}_M.
\end{equation}

As it is customary in supergravity, the gauging procedure is enforced by the introduction of prepotentials (or moment maps) for the gauged isometries.
In this context, the prepotential definition is 
\begin{equation}
	P_M^0 = -i k_M^i \partial_i K +i \, f_M,
\end{equation}
which, in the frame where a prepotential exists, becomes
\begin{equation}
	P_M^0 = e^K\, \overline{Z}^T \Omega T_M Z = e^K\,T_{MN}{}^Q \Omega_{QP} Z^N  \overline{Z}^P .
\end{equation}
Prepotentials satisfy the constraint
\begin{equation}\label{prepoconst}
	Z^M(z) P_M^0(z,\bar{z}) = 0
\end{equation}
which also implies 
\begin{equation}
	Z^M(z) k_M^I(z) = 0.
\end{equation}

After the gauging, the resulting scalar potential is
\begin{equation}\label{Dtermapp}
	{\cal V}_D = |\overline{V}^M k_M|^2 = \overline{V}^M k_M^I V^N \bar{k}_N^{\bar{J}} g_{I \bar J}.
\end{equation}


\section{Construction of the potential} 
\label{sec:construction_of_the_potential}

The potential in \cite{Fre:2002pd} and its generalization in \cite{Ogetbil:2008tk} have both an F-term and a D-term contribution.
For the F-term one constructs a U(1)$_R$ gauging by turning on a non-trivial charge $e_0$ in the direction 4 and rotating the sections with the symplectic matrix 
\begin{equation}
	S_1 = \left(\begin{array}{cccccc}
	{\mathbb 1}_4 & 0 &  & 0 &  & \\[2mm]
	 & \cos \delta & 0 &  & \sin \delta & \\[2mm]
	 & 0 & {\mathbb 1}_{p-1} & 0 &  &  \\[2mm]
	 & 0 &  & {\mathbb 1}_4 & 0 & \\
	 & -\sin \delta & 0 &  & \cos \delta & \\[2mm]
	 & 0 &  & 0 &  & {\mathbb 1}_{p-1}
	\end{array}\right)
\end{equation}
so that 
\begin{equation}
	W = \langle S_1 Z , Q\rangle = e_0 (\cos \delta\, X^4 + \sin \delta\, F_4).
\end{equation}

For the D-term one needs to identify the appropriate non-abelian group. 
To do so, first let us recall a few facts about the isometries of very special K\"ahler manifolds, i.e.~those with a cubic prepotential like in (\ref{cubicprep}).
From \cite{deWit:1991nm,deWit:1992wf}, the action of the isometries can be specified by a few parameters $\beta, a_I, b^I$ and a matrix $\tilde{B}^I{}_J$:
\begin{eqnarray}\label{deltaX0}
	\delta X^0 &=& \beta X^0 + a_I X^I, \\[2mm]
	\delta X^I &=& b^I X^0 + \frac13 \beta X^I + \tilde{B}^I{}_J X^J + A^{IJ} F_J, \\[2mm]
	\delta F_0 &=& - \beta F_0 - b^I F_I, \\[2mm]
	\label{deltaFI}
	\delta F_I &=& - a_I F_0 - \frac13 \beta F_I - F_J \tilde{B}^J{}_I + C_{IJK} X^J b^K.
\end{eqnarray}
In order to generate an isometry, the matrix $\widetilde{B}$ has to satisfy
\begin{equation}
	\widetilde{B}^I_{(J} C_{KL)I} = 0.
\end{equation}
The parameters $\beta$, $b^I$ and those in $\widetilde{B}$ are associated to perturbative symmetries, which do not mix $X^\Lambda$ and $F_{\Lambda}$ nor vectors with dual vectors, and are present for any such manifold.
Non-perturbative symmetries, mixing $X^\Lambda$ and $F_{\Lambda}$ (and hence vectors with dual vectors) are  parameterized by $a_K$ and exist only when
\begin{equation}
	A^{IJ} = {\rm e}^{2 K} C^{IJK} a_K
\end{equation}
is constant, for $a_K$ constants and $C^{IJK}$ obtained from $C_{IJK}$ by raising indices with the metric.
This happens for all symmetric spaces and therefore we focus once again on the $L(0,p)$ manifolds.

The isometries of the $L(0,p)$ manifolds are specified by (\ref{deltaX0})--(\ref{deltaFI}), where
\begin{equation}
	A^{IJ} = \left(\begin{array}{cccc}
	0 & a_2 & -a_3 & 0\\[2mm]
	a_2 & a_1 & 0 & -a_j\\[2mm]
	-a_3 & 0 & -a_1 & a_j\\[2mm]
	0 & -a_i & a_i & (a_3-a_2) \delta_{ij}
	\end{array}\right)
\end{equation}
and
\begin{equation}
	\widetilde{B} = \left(\begin{array}{ccccc}
	-2 \lambda & 0 & 0 & \alpha_j \\[2mm]
	0 & \lambda & \chi & \gamma_j \\[2mm]
	0 & \chi & \lambda &\gamma_j \\[2mm]
	\gamma_i & \frac{\gamma_i}{2} & \frac{\gamma_i}{2} & \frac{\chi-\lambda}{2}\,\delta_{ij} + \theta_{ij}
	\end{array}\right),
\end{equation}
where $\theta_{ji} = - \theta_{ij}$.

This means we can describe them by the $(2n_V+2)\times (2n_V+2)$ matrices
\begin{equation}
	T\left[\beta,b^I,a_I,\lambda,\chi,\alpha_i,\gamma_i,\theta_{ij}\right] = \left(\begin{array}{cccc}
	\beta & \vec{a} & 0 &  0 \\[2mm]
	\vec{b} & \frac{\beta}3  {\mathbb 1} + \widetilde{B} & 0 & A \\[2mm]
	0 & 0 & -\beta & - \vec{b} \\[2mm]
	0 & B & -\vec{a} & - \frac{\beta}3 {\mathbb 1} - \widetilde{B}^T 
	\end{array}\right),
\end{equation}
where $B_{IJ} = C_{IJK} b^k$.

Given the factorized structure of the scalar manifold we can identify 3 isometries as the SU(1,1) factor that commutes with all the rest.
These are generated by the parameter choices $a_2=-a_3=b_2=-b_3=\frac1{2\sqrt2}$, $b_2=-b_3=-a_2=a_3=\frac1{2\sqrt2}$ and $\beta = \frac12$, $\lambda = -\frac16$, $\chi = \frac12$, respectively, while setting everything else to zero.
We notice that such isometries act only on the scalar combination $s = z_3 - z_2$ and therefore we can find the other factor by looking at all isometries that leave $s$ invariant, namely all those where $a_2 = a_3$, $b_2 = b_3$ and $\beta = \frac32 (\lambda - \chi)$.
This leaves $(p+4)(p+3)/2$ generators among which we can select those generating an SO(2,1) that exists for any $p$ by considering 
\begin{eqnarray}
		T_1 &=& T\left[\chi = \frac12,\lambda=\frac12\right], \\[2mm] 
		T_2 &=& T\left[a_1=-a_2=-a_3=\frac1{\sqrt2},b_1=-2b_2=-2b_3=-\frac{1}{2\sqrt2}\right], \\[2mm]  
		T_3 &=& T\left[a_1=a_2=a_3=-\frac1{\sqrt2},b_1=2b_2=2b_3=\frac{1}{2\sqrt2}\right],
\end{eqnarray}
which satisfy $[T_1,T_2] = -T_3$, $[T_3,T_1] = T_2$ and $[T_2,T_3] = T_1$.
The action on the scalar fields is then
\begin{equation}
	\left\{
	\begin{array}{l}
	\displaystyle	\delta_1 z^1 = - z^1, \\[2mm]
	\displaystyle	\delta_1 (z^2+z^3) = \frac12 \left(z^2+z^3\right), \\[2mm]
	\displaystyle	\delta_1 z^i = 0, 
	\end{array}
	\right.
\end{equation}
\begin{equation}
	\left\{
	\begin{array}{l}
	\displaystyle	\delta_2 z^1 =  \frac1{2\sqrt2}\left(-1 + 2 (z^1)^2 - 2 z^i z^j \delta_{ij}\right), \\[2mm]
	\displaystyle	\delta_2 (z^2+z^3) = \frac{1}{4\sqrt2} \left(1- 2 (z^2+z^3)^2 + 2 \delta_{ij}z^i z^j\right), \\[2mm]
	\displaystyle	\delta_2 z^i = \frac1{\sqrt2}\, (z^1-z^2-z^3)z^i, 		
	\end{array}
	\right.
\end{equation}
\begin{equation}
	\left\{
	\begin{array}{l}
	\displaystyle 		\delta_3 z^1 = \frac1{2\sqrt2}\left(-1 + 2 (z^1)^2 - 2 z^i z^j \delta_{ij}\right), \\[2mm]
	\displaystyle 		\delta_3 (z^2+z^3) = \frac{1}{4\sqrt2} \left(1+ 2 (z^2+z^3)^2 + 2 \delta_{ij}z^i z^j\right), \\[2mm]
	\displaystyle 		\delta_3 z^i = \frac1{\sqrt2}\, (z^1+z^2+z^3)z^i. 
	\end{array}
	\right.
\end{equation}
As expected these isometries include the non-perturbative parameters $a_I$.

As we mentioned several times we can perform a symplectic rotation to a frame where such gauging becomes electric.
We can deduce the form of the symplectic matrix $S$ such that ${V}^\prime = S { V}$ and $ST_{1,2,3}S^{-1}$ becomes block-diagonal, by looking at the action of the generators mixing $X^\Lambda$ with $F_{\Lambda}$.
This does not completely fix $S$, but one consistent choice is 
\begin{equation}
	S = \left(\begin{array}{cccccccccccc}
	 0 & 0 & 0 & 0 & 1 & 0 & 0 & 0 & 0 & &0&0 \\
	 -\frac1{\sqrt2} & 0 & 0 & 0 & 0 & 0 & 0 & 0 & \sqrt2 &-\sqrt2  &0&0 \\
	 0 & 1 & 1 & 1 & 0 & 0 & 0 & 0 & 0 & 0 &0&0\\
	 0 & -1 & 1 & 1 & 0 & 0 & 0 & 0 & 0 & 0 &0&0\\
	  \frac1{\sqrt2} & 0 & 0 & 0 & 0 & 0 & 0 & 0 & \sqrt2  & -\sqrt2 &0&0 \\
	 0  & 0 & 0 & 0 & 0 & {\mathbb 1} & 0 & 0 & 0 & 0 &0&0 \\
	 0  & 0 & 0 & 0 & 0 & 0 & 0 & 0 & 0 & 0 &1& 0\\
	 0  & 0 & -\frac{1}{4\sqrt2} & \frac{1}{4\sqrt2} & 0 & 0 & -\frac1{\sqrt2} & 0 & 0 & 0 & 0 & 0\\
	 0  & 0 & 0 & 0 & 0 & 0 & 0 & \frac12 & \frac14 & \frac14 & 0 & 0\\
	 0  & 0 & 0 & 0 & 0 & 0 & 0 & -\frac12 & \frac14 & \frac14 & 0 & 0\\
	 0  & -\frac1{4\sqrt2} & \frac{1}{4\sqrt2} & \frac{1}{\sqrt2} & 0 & 0 & 0 & 0 & 0 & 0 & 0 & 0\\
	 0  & 0 & 0 & 0 & 0 & 0 & 0 & 0 & 0 & 0 & 0 & {\mathbb 1}
	\end{array}\right).
\end{equation}
In the new basis the generators act as simple rotations
\begin{eqnarray}
	&&\delta_1 \tilde{X}^3 = \tilde{X}^2, \ \delta_1 \tilde{X}^2 = \tilde{X}^3, \\[2mm]
	&&\delta_2 \tilde{X}^3 = -\tilde{X}^1, \ \delta_2 \tilde{X}^1 = -\tilde{X}^3, \\[2mm]
	&&\delta_3 \tilde{X}^3 = \tilde{X}^1, \ \delta_3 \tilde{X}^1 = -\tilde{X}^3.
\end{eqnarray}



\begin{thebibliography}{99}

\bibitem{Palti:2019pca} E.~Palti, \emph{``The Swampland: Introduction and Review,''}
Fortsch. Phys. \textbf{67} (2019) no.6, 1900037
[arXiv:1903.06239 [hep-th]].

\bibitem{ArkaniHamed:2006dz} N.~Arkani-Hamed, L.~Motl, A.~Nicolis and C.~Vafa, \emph{``The String landscape, black holes and gravity as the weakest force,''}
JHEP \textbf{06}, 060 (2007)
[arXiv:hep-th/0601001 [hep-th]].

\bibitem{Dine:1985he} M.~Dine and N.~Seiberg, \emph{``Is the Superstring Weakly Coupled?,''}
Phys. Lett. B \textbf{162} (1985), 299-302

\bibitem{Montero:2019ekk} M.~Montero, T.~Van Riet and G.~Venken, \emph{``Festina Lente: EFT Constraints from Charged Black Hole Evaporation in de Sitter,''}
JHEP \textbf{01} (2020), 039
[arXiv:1910.01648 [hep-th]].

\bibitem{Antoniadis:2020xso} I.~Antoniadis and K.~Benakli, \emph{``Weak Gravity Conjecture in de Sitter space-time,''}
Fortsch. Phys. \textbf{68} (2020) no.9, 2000054
[arXiv:2006.12512 [hep-th]].

\bibitem{Huang:2006hc} Q.~G.~Huang, M.~Li and W.~Song, \emph{``Weak gravity conjecture in the asymptotical dS and AdS background,''}
JHEP \textbf{10} (2006), 059
[arXiv:hep-th/0603127 [hep-th]].

\bibitem{Carroll:2004st} S.~M.~Carroll, \emph{``Spacetime and Geometry,'' }
Addison-Wesley (2004).


\bibitem{Cribiori:2020wch} N.~Cribiori, F.~Farakos and G.~Tringas, \emph{``Three-forms and Fayet-Iliopoulos terms in Supergravity: Scanning Planck mass and BPS domain walls,''}
JHEP \textbf{05} (2020), 060
[arXiv:2001.05757 [hep-th]].

\bibitem{Fre:2002pd} P.~Fre, M.~Trigiante and A.~Van Proeyen, \emph{``Stable de Sitter vacua from N=2 supergravity,''}
Class. Quant. Grav. \textbf{19} (2002), 4167-4194
[arXiv:hep-th/0205119 [hep-th]].

\bibitem{deWit:1984wbb} B.~de Wit and A.~Van Proeyen, \emph{``Potentials and Symmetries of General Gauged N=2 Supergravity: Yang-Mills Models,''}
Nucl. Phys. B \textbf{245} (1984), 89-117

\bibitem{deWit:1991nm} B.~de Wit and A.~Van Proeyen, ``Special geometry, cubic polynomials and homogeneous quaternionic spaces,''
Commun. Math. Phys. \textbf{149} (1992), 307-334
[arXiv:hep-th/9112027 [hep-th]].

\bibitem{deWit:1992wf} B.~de Wit, F.~Vanderseypen and A.~Van Proeyen, \emph{``Symmetry structure of special geometries,''}
Nucl. Phys. B \textbf{400} (1993), 463-524
[arXiv:hep-th/9210068 [hep-th]].

\bibitem{Ogetbil:2008tk} O.~Ogetbil, \emph{``Stable de Sitter Vacua in 4 Dimensional Supergravity Originating from 5 Dimensions,''}
Phys. Rev. D \textbf{78} (2008), 105001
[arXiv:0809.0544 [hep-th]].

\bibitem{Roest:2009tt} D.~Roest and J.~Rosseel, \emph{``De Sitter in Extended Supergravity,''}
Phys. Lett. B \textbf{685} (2010), 201-207
[arXiv:0912.4440 [hep-th]].

\bibitem{Borghese:2011en} A.~Borghese, R.~Linares and D.~Roest, \emph{``Minimal Stability in Maximal Supergravity,''}
JHEP \textbf{07} (2012), 034
[arXiv:1112.3939 [hep-th]].

\bibitem{DallAgata:2012plb} G.~Dall'Agata and G.~Inverso, \emph{``de Sitter vacua in N = 8 supergravity and slow-roll conditions,''}
Phys. Lett. B \textbf{718} (2013), 1132-1136
[arXiv:1211.3414 [hep-th]].

\bibitem{Catino:2013syn} F.~Catino, C.~A.~Scrucca and P.~Smyth, \emph{``Simple metastable de Sitter vacua in N=2 gauged supergravity,''}
JHEP \textbf{04} (2013), 056
[arXiv:1302.1754 [hep-th]].

\bibitem{Obied:2018sgi} G.~Obied, H.~Ooguri, L.~Spodyneiko and C.~Vafa, \emph{``De Sitter Space and the Swampland,''}
[arXiv:1806.08362 [hep-th]].

\bibitem{Denef:2018etk}
F.~Denef, A.~Hebecker and T.~Wrase,
\emph{``de Sitter swampland conjecture and the Higgs potential,''}
Phys. Rev. D \textbf{98} (2018) no.8, 086004
[arXiv:1807.06581 [hep-th]].

\bibitem{Ooguri:2018wrx}
H.~Ooguri, E.~Palti, G.~Shiu and C.~Vafa,
\emph{``Distance and de Sitter Conjectures on the Swampland,''}
Phys. Lett. B \textbf{788} (2019), 180-184
[arXiv:1810.05506 [hep-th]].

\bibitem{Andriot:2018wzk}
D.~Andriot,
\emph{``On the de Sitter swampland criterion,''}
Phys. Lett. B \textbf{785} (2018), 570-573
[arXiv:1806.10999 [hep-th]].

\bibitem{Garg:2018reu}
S.~K.~Garg and C.~Krishnan,
\emph{``Bounds on Slow Roll and the de Sitter Swampland,''}
JHEP \textbf{11} (2019), 075
[arXiv:1807.05193 [hep-th]].

\bibitem{Andriot:2018mav}
D.~Andriot and C.~Roupec,
\emph{``Further refining the de Sitter swampland conjecture,''}
Fortsch. Phys. \textbf{67} (2019) no.1-2, 1800105
[arXiv:1811.08889 [hep-th]].



\bibitem{Rudelius:2019cfh}
T.~Rudelius,
\emph{``Conditions for (No) Eternal Inflation,''}
JCAP \textbf{08} (2019), 009
[arXiv:1905.05198 [hep-th]].

\bibitem{Andrianopoli:1996cm} L.~Andrianopoli, M.~Bertolini, A.~Ceresole, R.~D'Auria, S.~Ferrara, P.~Fre and T.~Magri,
\emph{``N=2 supergravity and N=2 superYang-Mills theory on general scalar manifolds: Symplectic covariance, gaugings and the momentum map,''} J. Geom. Phys. \textbf{23} (1997), 111-189
[arXiv:hep-th/9605032 [hep-th]].

\bibitem{DallAgata:2011aa} G.~Dall'Agata and G.~Inverso, \emph{``On the Vacua of N = 8 Gauged Supergravity in 4 Dimensions,''}
Nucl. Phys. B \textbf{859} (2012), 70-95
[arXiv:1112.3345 [hep-th]].

\bibitem{DallAgata:2005zlf} G.~Dall'Agata and N.~Prezas, \emph{``Scherk-Schwarz reduction of M-theory on G2-manifolds with fluxes,''} JHEP \textbf{10} (2005), 103 [arXiv:hep-th/0509052 [hep-th]].

\bibitem{DallAgata:2012mfj} G.~Dall'Agata, G.~Inverso and M.~Trigiante, \emph{``Evidence for a family of SO(8) gauged supergravity theories,''} Phys. Rev. Lett. \textbf{109} (2012), 201301 [arXiv:1209.0760 [hep-th]].

\bibitem{Strominger:1990pd} A.~Strominger, ``SPECIAL GEOMETRY,'' Commun. Math. Phys. \textbf{133} (1990), 163-180

\bibitem{Ceresole:1995jg} A.~Ceresole, R.~D'Auria, S.~Ferrara and A.~Van Proeyen, \emph{``Duality transformations in supersymmetric Yang-Mills theories coupled to supergravity,''} Nucl. Phys. B \textbf{444} (1995), 92-124
[arXiv:hep-th/9502072 [hep-th]].

\bibitem{Ceresole:1995ca} A.~Ceresole, R.~D'Auria and S.~Ferrara, \emph{``The Symplectic structure of N=2 supergravity and its central extension,''}
Nucl. Phys. B Proc. Suppl. \textbf{46} (1996), 67-74
doi:10.1016/0920-5632(96)00008-4
[arXiv:hep-th/9509160 [hep-th]].

\bibitem{DallAgata:2003sjo} G.~Dall'Agata, R.~D'Auria, L.~Sommovigo and S.~Vaula, \emph{``D = 4, N=2 gauged supergravity in the presence of tensor multiplets,''}
Nucl. Phys. B \textbf{682} (2004), 243-264
[arXiv:hep-th/0312210 [hep-th]].

\bibitem{DAuria:2004yjt} R.~D'Auria, L.~Sommovigo and S.~Vaula, \emph{``N = 2 supergravity Lagrangian coupled to tensor multiplets with electric and magnetic fluxes,''}
JHEP \textbf{11} (2004), 028
[arXiv:hep-th/0409097 [hep-th]].

\bibitem{Andrianopoli:2007ep} L.~Andrianopoli, R.~D'Auria and L.~Sommovigo, \emph{``D=4, N=2 supergravity in the presence of vector-tensor multiplets and the role of higher p-forms in the framework of free differential algebras,''} 
Adv. Stud. Theor. Phys. \textbf{1} (2008), 561-596
[arXiv:0710.3107 [hep-th]].

\bibitem{DallAgata:2010ejj} G.~Dall'Agata and A.~Gnecchi, \emph{``Flow equations and attractors for black holes in N = 2 U(1) gauged supergravity,''}
JHEP \textbf{03} (2011), 037
[arXiv:1012.3756 [hep-th]].

\bibitem{Louis:2009xd} J.~Louis, P.~Smyth and H.~Triendl, \emph{``Spontaneous N=2 to N=1 Supersymmetry Breaking in Supergravity and Type II String Theory,''}
JHEP \textbf{02} (2010), 103
[arXiv:0911.5077 [hep-th]].

\bibitem{deWit:2011gk} B.~de Wit and M.~van Zalk, \emph{``Electric and magnetic charges in N=2 conformal supergravity theories,''}
JHEP \textbf{10} (2011), 050
[arXiv:1107.3305 [hep-th]].










\end{thebibliography}
\end{document}